\documentclass[a4paper,12pt]{article}%
\usepackage{cite}
\usepackage{amsmath}
\usepackage{amsfonts}
\usepackage{amssymb}
\usepackage{graphicx}%
\providecommand{\U}[1]{\protect\rule{.1in}{.1in}}
\begin{document}

\title{Irradiation Induced Dimensional Changes in Bulk Graphite; The theory}
\author{S.V. Panyukov\\
\emph{\small P.N. Lebedev Physics Institute,
Russian Academy of Sciences, Moscow, 117924, Russia}\\
A.V. Subbotin and M.V. Arjakov\\
\emph{\small Scientific and Production Complex Atomtekhnoprom, Moscow, 119180,
Russia}}

\begin{abstract}
Basing on experimental data on irradiation-induced deformation of graphite we
introduced a concept of diffuse domain structure developed in reactor graphite
produced by extrusion. Such domains are considered as random continuous
deviations of local graphite texture from the global one. We elucidate the
origin of domain structure and estimate the size and the degree of
orientational ordering of its domains. Using this concept we explain the well
known radiation-induced size effect observed in reactor graphite. We also
propose a method for converting the experimental data on shape-change of
finite-size samples to bulk graphite. This method gives a more accurate
evaluation of corresponding data used in estimations of reactor graphite
components lifetime under irradiation.

\end{abstract}
\maketitle
\tableofcontents

\section{Introduction}

Employment of graphite as neutron moderator as well as construction material
has a long history starting with the first reactor created by Enrico Fermi.
Artificial reactor grade graphites were designed for these purposes\cite{1n}.
The graphite components life time in active zone actually determines a life
time of water-cooled carbon reactor as a whole, stimulating a study of
radiation-induced evolution of reactor graphite properties.

With respect to production technology, reactor grade graphites range from
almost isotropic to substantially anisotropic. Here we imply both physical and
mechanical properties as well as irradiation effects. Widely used reactor
grade graphites, such as Pile Grade A, ATR-2E, GR-280, produced by extrusion
process exhibit a transversal-isotropic symmetry of physical and mechanical
properties and radiation-induced effects\cite{2n,3n,4n,5n,6n,7n}. Such
symmetry requires five independent mechanical constants; in different
representations these are elasticity constants, or compliances, or the set of
Young's modulus, rigidity modulus, and Poisson coefficient\cite{21n}.

Graphite rod subjected to irradiation undergoes transversely isotropic
alternating-sign shape changes reaching several percent. The most important
factors leading to non-uniform shape changes and hence, growth in internal
stresses, are gradients of temperature and neutron exposure on the scale of a
particular monolithic graphite element. Radiation-induced creep of graphite is
not sufficient to compensate effects caused by shape changes and temperature
gradients. Finally, a complicated irradiation-induced evolution of elastic
moduli along with shape changes may lead, under low creep, to a destruction of
monolithic graphite components.

In calculations of construction lifetime by using, e.g., the crack initiation
and growth criterion, the determining factors are the data on evolution with
neutron fluence of elastic moduli and relative shape changes in directions of
an axis of symmetry ($^{\parallel}$) and in a plane of isotropy ($^{\perp}%
$)\cite{2n,3n,4n,5n,6n,7n,8n,9n,10n,11n}. The standard way to obtain such data
is by irradiation of finite-size graphite samples in materials testing
reactors at a higher, compared with power reactors, neutron flux accompanied
with periodical measurements of characteristics. In 5-6 years the neutron
fluence is acquired corresponding to 30-40 years lifetime graphite unit in
operation conditions of power reactor. Such experiments are inevitably
conducted under conditions different from operation ones, which has an
inevitable influence on the data obtained. Necessary adjustments require a
development of the corresponding models which would help to convert the data
obtained under higher neutron flux for finite-dimension samples to conditions
of bulk graphite and power reactors.

\section{Formulation of the problem}

In adjusting data on transversally isotropic reactor grade graphite irradiated
in materials testing reactors one encounters three main problems, requiring a
creation of models and a recalculation of corresponding data:

\begin{enumerate}
\item[\textbf{I.}] \textbf{Differences in neutron energy spectra. }The problem
is sufficiently simply solved by standard methods of neutron fluence
recalculation to solid-state doze values in terms of displacements per atom
($\operatorname{dpa}$) \cite{5n,n+7}. Note that such conversion is not
universal. When heterogeneous nucleation of microstructural elements is
important (for example, PWR pressure vessels) the conversion to dpa results in
a partial loss of needed information.

\item[\textbf{II.}] \textbf{Noticeable differences in intensities of Frenkel
pair generation (}$\operatorname{dpa}/\operatorname*{s}$\textbf{).} Morphology
of reactor grade graphite has a complicated multiscale structure formed by the
filler, which comprises the objects with different degrees of order based on
graphite micro-crystallites with the ideal crystal structure and dimensions of
$\sim10^{2}\operatorname{nm}$, a binder with an isotropic microcrystalline
structure, and an ensemble of microcracks\cite{1n,6n,8n,9n,10n,11n,12n,13n}.
By the objects are meant formations of various scales from micro-crystallite
complexes with various packing to larger formations including grains. As the
dimension scale increases, the objects become less anisotropic. An ensemble of
microcracks is a result of relaxation processes at the scales with
sufficiently high anisotropy level. At larger scales the anisotropy is too
small to initiate the microcracks initiation.

A driving force of radiation-induced effects in reactor grade graphite are
processes taking place in an ideal crystalline structure of microcrystallite.
Frenkel pair generation under the action of a neutron flux on crystal lattice
results in the formation of a supersaturated two-component solid solution of
interstitial atoms and vacancies. For a variety of reasons, the decay of
quasi-2D supersaturated solid solution mainly leads to the formation of
ensembles of interstitial dislocation loops of basal type, which, in turn,
change microcrystallite shapes\cite{6n,7n,11n,15n}. Below we list current
understanding of the microstructural mechanisms of radiation-induced shape
changes of microcrystallites: The formation of interstitial clusters,
dislocation loops and new graphitic planes will cause an expansion of the
microcrystallite in the c-axis\cite{13a}. Adjacent lattice vacancies collapse
parallel to the layers and form sinks for other vacancies, causing shrinkage
parallel to the graphite layers\cite{15a}{\LARGE .} Just the change of
crystallite shape is a driving force for developing stresses, evolution of
crack ensembles, and so on. At the bulk graphite scale, these effects are
responsible for the shape changes, evolution of mechanical properties
(partially) etc. Thus, juxtaposing radiation-induced effects obtained under
different generation rates of Frenkel pairs ($\operatorname{dpa}%
/\operatorname*{s}$) requires solving the nontrivial problem on development of
quasi-2D microstructure in graphite crystal lattice with microcrystal
self-consistent boundary conditions\cite{14n}.

\item[\textbf{III.}] \textbf{Discrepancy between radiation-induced shape
changes of bulk graphite and data obtained on finite-dimension samples.} In
our previous work \cite{16n}, based on numerous measurements of shape changes
in cylindrical samples the two results were obtained:

Initially circular cross-section of the sample under irradiation takes an
increasingly pronounced elliptical shape and orientations of the ellipses vary
randomly along the sample on a scale of $0.6\operatorname{cm}$. The results of
calculations of the relative volume changes using traditionally accepted
methodology
\begin{equation}
\left(  \Delta V/V\right)  _{T}=\left(  \Delta L/L\right)  ^{\parallel
}+2\left(  \Delta L/L\right)  ^{\perp}, \label{dV/V}%
\end{equation}
(where $\left(  \Delta L/L\right)  ^{\parallel}$ and $\left(  \Delta
L/L\right)  ^{\perp}$ are relative length changes of the samples cut from the
bulk material in directions parallel and perpendicular to the extrusion
direction) increasingly diverge with irradiation doze from the results of the
direct measurements of the relative volume changes (see Fig.~\ref{VOLUME}).
\begin{figure}
[tbh]
\begin{center}
\includegraphics[
height=2.2886in,
width=2.3311in
]%
{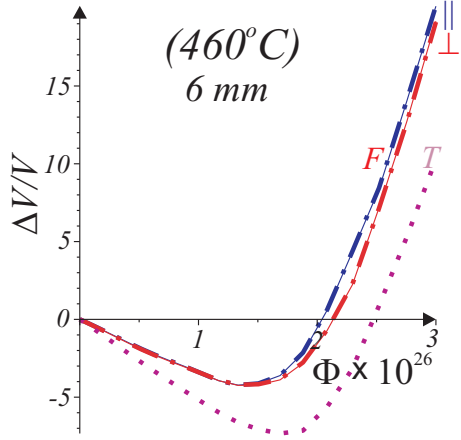}%
\caption{Relative volume change $\Delta V/V$ (in \%) as a function of fluence
$\Phi$ ($\operatorname{neutrons}/\operatorname{m}^{2}$, $E_{n}%
>0.18\operatorname{MeV}$) in samples irradiated at temperature $460\pm
25^{\circ}\operatorname{C}$. Two dash-dotted lines are experimentally measured
volumes $\Delta L/L+2\left(  \Delta d/d\right)  $ in $d=6\operatorname{mm}$
diameter cylindrical samples of length $L$ with both parallel ($^{\parallel}$)
and perpendicular ($^{\perp}$) cuts relative to the extrusion direction;
single dotted line ($T$) shows the traditionally used combination~(\ref{dV/V}%
), calculated using longitudinal experimental data for those two extrusions.
For the sample with $d=8\operatorname{mm}$, the result is qualitatively the
same.}%
\label{VOLUME}%
\end{center}
\end{figure}
As it was shown\cite{16n} the reason for such a mismatch is that
expression~(\ref{dV/V}) only relates the relative elongations of samples to
relative volume change while neglecting changes in cross-sectional areas.
\end{enumerate}

Basing on the presented experimental facts it was assumed that at macro-scale
graphite can be considered as an transversally-isotropic elastic medium with
irradiation-induced shape changes described by gradient tensor:%
\begin{equation}
\hat{F}_{M}=\left(
\begin{array}
[c]{ccc}%
1+\left(  \Delta L/L\right)  ^{\perp} & 0 & 0\\
0 & 1+\left(  \Delta L/L\right)  ^{\perp} & 0\\
0 & 0 & 1+\left(  \Delta L/L\right)  ^{\parallel}%
\end{array}
\right)  . \label{FM}%
\end{equation}
Here, $\left(  \Delta L/L\right)  ^{\parallel,\perp}$ are the relative bulk
graphite shape changes in parallel and perpendicular directions relative to
the axis of symmetry (which coincides with the extrusion direction). But on a
scale of $\simeq0.6\operatorname{cm}$, nonuniformities in shape changes are observed.

\begin{figure}
[tbh]
\begin{center}
\includegraphics[
height=2.1955in,
width=2.3993in
]%
{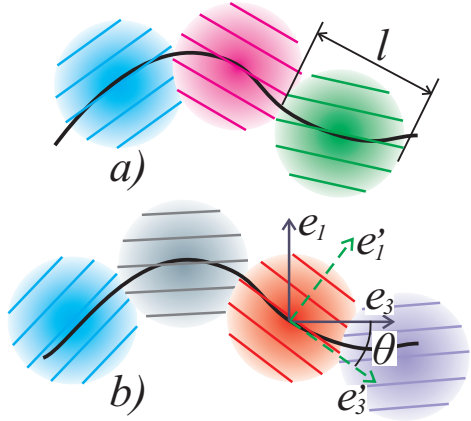}%
\caption{The random elastic medium with direction of anisotropy (solid line)
continuously varying in space on the scale $l$ can be visualized in several
equivalent ways a) and b) as the sets of diffuse domains of size $l$. The axis
$\mathbf{e}_{3}^{\prime}$ (anisotropy direction) of local coordinate system
$\left(  \mathbf{e}_{1}^{\prime},\mathbf{e}_{2}^{\prime},\mathbf{e}%
_{3}^{\prime}\right)  $ related to domains is rotated by random angle
$\theta=\theta\left(  \mathbf{x}\right)  $ with respect to the axis
$\mathbf{e}_{3}$ (orthotropy direction) of global coordinate system $\left(
\mathbf{e}_{1},\mathbf{e}_{2},\mathbf{e}_{3}\right)  $ related to the sample.}%
\label{DIFFUSE}%
\end{center}
\end{figure}

Actually, to such regions without pronounced boundaries may correspond local
(at a scale of $l\simeq0.6\operatorname{cm}$) random deviations of the axis of
symmetry from the average direction in the bulk graphite. This representation
was formalized by introducing a concept of \textquotedblleft
domain\textquotedblright\ -- the region of size $l$ possessing the gradient
tensor of relative radiation-induced deformation in local coordinate system%
\begin{equation}
\hat{F}_{0}=\left(
\begin{array}
[c]{ccc}%
1+\left(  \Delta l/l\right)  _{\perp} & 0 & 0\\
0 & 1+\left(  \Delta l/l\right)  _{\perp} & 0\\
0 & 0 & 1+\left(  \Delta l/l\right)  _{\parallel}%
\end{array}
\right)  . \label{F0}%
\end{equation}
Here $\left(  \Delta l/l\right)  _{\alpha}$ are irradiation-induced relative
changes in sizes of free domain (i.e. mentially cut out of the elastic
medium). The axis of symmetry of the domain is inclined at random angle
$\theta$ with respect to that of bulk graphite (see Fig.~\ref{DIFFUSE}).%
The domains are continuously transformed in space into each other on the scale
of domain size $l$. It is impossible to uniquely identify the boundaries of
such \textquotedblleft diffuse\textquotedblright\ domains, which can be
defined only statistically as regions in space with correlated directions of anisotropy.

The longitudinal measurement data $\left(  \Delta L/L\right)  ^{\parallel
},\left(  \Delta L/L\right)  ^{\perp}$ can be calculated by averaging over
random angle $\theta$ the projections of tensors $\hat{F}_{0}$ to the axis of
symmetry of cylindrical samples. Such formalism leads us to writing a closed
system of equations for obtaining $\left(  \Delta l/l\right)  _{\parallel
},\left(  \Delta l/l\right)  _{\perp}$ and $\overline{\sin^{2}\theta}$, where
$\theta$ is the angular deviation of the domain axis of symmetry from the bulk
graphite common axis of symmetry, by using experimentally measured parameters
$\left(  \Delta L/L\right)  _{S}^{\parallel},\left(  \Delta L/L\right)
_{S}^{\perp}$, and $\bar{\varepsilon}=\frac{1}{N}\sum\nolimits_{k=1}%
^{N}\varepsilon_{k}$ where $\varepsilon_{k}=\left[  d_{\max}-\left(  d_{\min
}\right)  _{k}\right]  /d$ is the flattening factor for $k$-th transversal
cross-section of the sample.

The developed model consistently explains the entire array of obtained
experimental data. Taking into account that sample cross-sections are of the
same scale ($d=0.6\operatorname{cm}$ and $d=0.8\operatorname{cm}$) as domain
size, the irradiation-induced change of domain shape in samples can be
considered as almost free dilatation (parameters $\left(  \Delta l/l\right)
_{\parallel}$ and $\left(  \Delta l/l\right)  _{\perp}$). At the same time, if
we consider the case of radiation-induced effects in bulk graphite, the
presence of an ensemble of disoriented domains with changing shapes would give
rise to internal stresses, which entails a conclusion that shape and volume
changes in bulk graphite are not equivalent to those in finite-size
cylindrical samples. Keeping in mind that we are aimed at determining shape
changes of bulk graphite while experimental data refer to samples with
cross-sections on the same scale as the domains size, we have to recalculate
the experimental data.

This work is devoted to further development of model \cite{16n} in order to
suggest a method for recalculating data obtained on small-size samples to
radiation-induced shape changes of bulk graphite. In the model developed for
transversal-isotropic medium we employ pure elastic approach rather than
elastoplastic one as required for rigorous solution of the problem. This is
explained by insufficient understanding of the mechanisms of graphite
irradiation creep related to the problem II above. Actually we obtain a lower
bound for influence of disoriented domain structure on shape changes in bulk graphite.

\begin{figure}
[tbh]
\begin{center}
\includegraphics[
height=2.7812in,
width=4.0524in
]%
{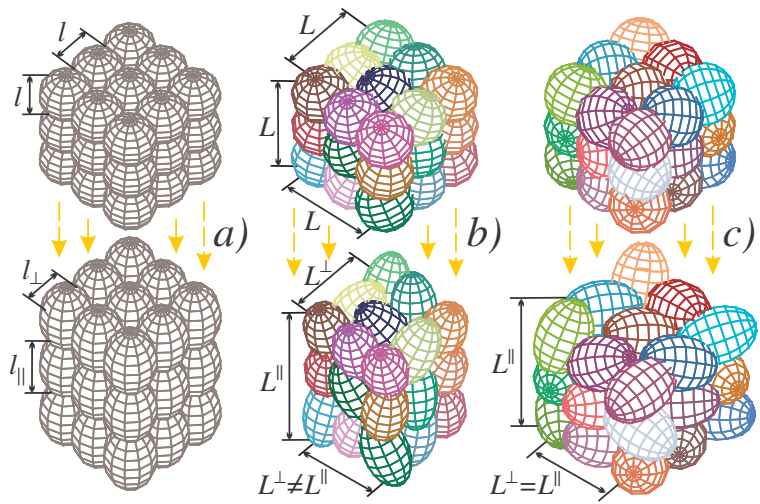}%
\caption{Domain structures in graphite: spherical shape of diffuse domains
with transversely isotropic symmetry (shown by grid lines) and size $l$
becomes ellipsoidal (with sizes $l_{\perp}\neq l_{\parallel}$) due to
irradiation-induced shape-change of free domains. Regular structure a) deforms
affinely ($L^{\perp}/L^{\parallel}=l_{\perp}/l_{\parallel}$). The domains in
irregular domain structure b) and c) randomly deform due to interaction with
their elastic environment, and the structure as the whole deforms nonaffinelly
($L^{\perp}/L^{\parallel}\neq l_{\perp}/l_{\parallel}$): In case b) of high
anisotropy the macroscopic deformation has the same transversely isotropic
symmetry ($L^{\perp}\neq L^{\parallel}$), but different amplitude compared
with the domains. In case c) of isotropic distribution of domain directions
the macroscopic deformation is isotropic ($L^{\perp}=L^{\parallel}$) even for
anisotropic domain deformation ($l_{\perp}\neq l_{\parallel}$).}%
\label{fig1}%
\end{center}
\end{figure}

In section~\ref{ELASTICITY} we develop a new approach to the description of
elasticity of such diffuse domain structure. We consider both the case of high
anisotropy (see Fig.~\ref{fig1} b) and uniform distribution of anisotropy
directions (see Fig.~\ref{fig1} c).
Those cases correspond to different grades of reactor
graphite\cite{1n,4n,5n,6n}. Macroscopic description of diffuse domain
structure on scales large with respect to domain size is developed in
section~\ref{MACRO}. In section~\ref{FORMATION} we describe formation of
diffuse domain structure in graphite produced by the extrusion process. We
also estimate size and a degree of alignment of the emerging domains.

\section{Elasticity of diffuse domain structure\label{ELASTICITY}}

In this section we introduce main concepts and formulate a physical model of
elasticity of diffuse domain structure. We consider diffuse domains as more or
less homogeneous regions with anisotropic elasticity. Such domains do not have
sharp boundaries, and their direction of anisotropy continuously and randomly
change in space on scale of domain size $l$, see Fig.~\ref{DIFFUSE}.

\subsection{Strain of domain structure}

Deformation of solid can be described by strain tensor $\hat{\varepsilon}$,
which is related to the gradient tensor as $\hat{\varepsilon}=(\hat{F}\hat
{F}^{T}-1)/2$, where $^{T}$ means transposition. Since the domain size is much
larger than the scale of micro-cracks the strain tensor can be considered as a
continuous function of coordinates ${\mathbf{x}}$. We describe deformation of
graphite by taking the state before irradiation as reference state. Thus
defined strain tensor can be presented as the sum $\hat{\varepsilon}%
^{tot}=\hat{\varepsilon}^{M}+\hat{\varepsilon}$ of two contributions:
macroscopic deformation of sample $\hat{\varepsilon}^{M}\left(  {\mathbf{x}%
}\right)  $ and local elastic deformation $\hat{\varepsilon}\left(
{\mathbf{x}}\right)  $ with components%
\begin{equation}
\varepsilon_{\alpha\beta}=\frac{1}{2}\left(  \frac{\partial u_{\alpha}%
}{\partial x_{\beta}}+\frac{\partial u_{\beta}}{\partial x_{\alpha}}\right)
\label{eab}%
\end{equation}
where $u_{\alpha}$ are components of elastic deformation vector. In
equilibrium, this vector describes local deviations on the scale of domain
size $l$, so that its average over the macro-scale vanishes.

In the case of bulk graphite with free boundaries the macroscopic strain
tensor $\hat{\varepsilon}^{M}$ is diagonal in global coordinate system related
to sample (see Fig.~\ref{DIFFUSE}). In linear approximation in relative
changes of sample sizes $\left(  \Delta L/L\right)  _{\alpha}$ in principal
directions $\alpha=1,2,3$ (see Fig.~\ref{fig1} b) it can be written as%
\begin{equation}
\varepsilon_{\alpha\beta}^{M}=\left(  F_{M}\right)  _{\alpha\beta}%
-\delta_{\alpha\beta}=\left(  \Delta L/L\right)  _{\alpha}\delta_{\alpha\beta}
\label{eir}%
\end{equation}
Here and below, if possible, we shall drop the index $M$ characterizing the
bulk graphite. Gradient tensor $\hat{F}_{M}$ is defined in Eq.~(\ref{FM}) and
$\delta$ is the unit Kronecker tensor.

Graphite domains are randomly oriented in space. The local coordinate system
related to the domains (see Fig.~\ref{DIFFUSE} b) can be obtained from global
coordinate system related to sample by means of rotations by three Euler
angles $\Omega$: two polar angles $\varphi$ and $\psi$ and the azimuthal angle
$\theta$, characterizing the orientation of the local frame with respect to
the graphite orthotropy axis. All three angles $\Omega\left(  {\mathbf{x}%
}\right)  =\left(  \varphi\left(  {\mathbf{x}}\right)  ,\theta\left(
{\mathbf{x}}\right)  ,\psi\left(  {\mathbf{x}}\right)  \right)  $ randomly
vary in space with coordinates ${\mathbf{x}}$ on the scale of domain size $l$.
Domain rotation by given angles $\Omega=\left(  \varphi,\theta,\psi\right)  $
can be described by rotation matrix:
\begin{equation}%
\begin{array}
[c]{c}%
\hat{r}\left(  \Omega\right)  =\left(
\begin{array}
[c]{ccc}%
\cos\varphi & -\sin\varphi & 0\\
\sin\varphi & \cos\varphi & 0\\
0 & 0 & 1
\end{array}
\right)  \times\\
\left(
\begin{array}
[c]{ccc}%
1 & 0 & 0\\
0 & \cos\theta & -\sin\theta\\
0 & \sin\theta & \cos\theta
\end{array}
\right)  \left(
\begin{array}
[c]{ccc}%
\cos\psi & -\sin\psi & 0\\
\sin\psi & \cos\psi & 0\\
0 & 0 & 1
\end{array}
\right)
\end{array}
\label{R}%
\end{equation}
This matrix relates the strain tensors in global $\hat{\varepsilon}^{tot}%
=\hat{\varepsilon}^{M}+\hat{\varepsilon}$ and in local $\hat{\varepsilon
}^{loc}$ coordinate systems:
\begin{equation}
\varepsilon_{\alpha\beta}^{tot}=\sum\nolimits_{\alpha^{\prime}\beta^{\prime}%
}r_{\alpha\alpha^{\prime}}r_{\beta\beta^{\prime}}\varepsilon_{\alpha^{\prime
}\beta^{\prime}}^{loc} \label{eRR}%
\end{equation}
Eq.~(\ref{eRR}) can be rewritten in matrix notations as
\begin{equation}
\varepsilon_{i}^{tot}=\sum\nolimits_{j}R_{ij}\left(  \Omega\right)
\varepsilon_{j}^{loc} \label{matrix}%
\end{equation}

Here and below we use Greek indexes ($\alpha\beta$) to distinguish components
of symmetric tensor $\varepsilon_{\alpha\beta}=\varepsilon_{\beta\alpha}$ and
Latin indexes ($i$) when presenting this tensor in vector notation. The first
three components of such vector determine normal strain $\varepsilon
_{i}=\varepsilon_{ii}$ ($i=1,2,3$) and the other three components of this
vector determine shear strain, $\varepsilon_{4}=2\varepsilon_{23}%
,\varepsilon_{5}=2\varepsilon_{13}$ and $\varepsilon_{6}=2\varepsilon_{12}$.
Similar expressions for stress tensor are $\sigma_{i}=\sigma_{ii}$ ($i=1,2,3$)
and $\sigma_{4}=\sigma_{23},\sigma_{5}=\sigma_{13},\sigma_{6}=\sigma_{12}$.

The rotation matrix $\hat{R}\left(  \Omega\right)  $ in Eq.~(\ref{matrix}) is
\[%
\begin{array}
[c]{c}%
\hat{R}\left(  \Omega\right)  =\left(
\begin{array}
[c]{ccc}%
r_{11}^{2} & r_{12}^{2} & r_{13}^{2}\\
r_{21}^{2} & r_{22}^{2} & r_{23}^{2}\\
r_{31}^{2} & r_{32}^{2} & r_{33}^{2}\\
2r_{21}r_{31} & 2r_{22}r_{32} & 2r_{23}r_{33}\\
2r_{11}r_{31} & 2r_{12}r_{32} & 2r_{13}r_{33}\\
2r_{11}r_{21} & 2r_{12}r_{22} & 2r_{13}r_{23}%
\end{array}
\right. \\
\left.
\begin{array}
[c]{ccc}%
r_{13}r_{12} & r_{13}r_{11} & r_{11}r_{12}\\
r_{23}r_{22} & r_{23}r_{21} & r_{21}r_{22}\\
r_{33}r_{32} & r_{33}r_{31} & r_{31}r_{32}\\
r_{23}r_{32}+r_{22}r_{33} & r_{23}r_{31}+r_{21}r_{33} & r_{22}r_{31}%
+r_{21}r_{32}\\
r_{11}r_{32}+r_{11}r_{32} & r_{13}r_{31}+r_{11}r_{33} & r_{12}r_{31}%
+r_{11}r_{32}\\
r_{11}r_{22}+r_{12}r_{23} & r_{13}r_{21}+r_{11}r_{23} & r_{12}r_{21}%
+r_{11}r_{22}%
\end{array}
\right)
\end{array}
\]
and where $r_{ij}$ are elements of the matrix~(\ref{R}). Below we use
notations $\hat{R}^{-1}\left(  \Omega\right)  ,\hat{R}^{T}\left(
\Omega\right)  $ and $\hat{R}^{-T}\left(  \Omega\right)  $ for inverse,
transposed and inverse transposed rotation matrixes, respectively.

\subsection{Elastic energy of domain structure}

Energy of diffuse domain structure takes simplest form in local coordinate
system, which continuously rotates in space (see Fig.~\ref{DIFFUSE} b):
\begin{equation}
E=\dfrac{1}{2}\int d{\mathbf{x}}{%
{\displaystyle\sum\nolimits_{ij}}
}c_{ij}\left[  \hat{\varepsilon}^{loc}\left(  {\mathbf{x}}\right)
-\hat{\varepsilon}^{F}\right]  _{i}\left[  \hat{\varepsilon}^{loc}\left(
{\mathbf{x}}\right)  -\hat{\varepsilon}^{F}\right]  _{j} \label{E1}%
\end{equation}
Irradiation-induced dilatation of freely dilating domains $\hat{\varepsilon
}^{F}$ describes irreversible deformation of the domain with free boundaries
(which was cut out of the sample) due to irradiation. The dilatation is
diagonal in local coordinate system related to the domain:
\begin{equation}
\varepsilon_{\alpha\beta}^{F}=\left(  \hat{F}_{0}\right)  _{\alpha\beta
}-\delta_{\alpha\beta}=\left(  \Delta l/l\right)  _{\alpha}\delta_{\alpha
\beta} \label{e_free}%
\end{equation}
Here $\hat{F}_{0}$ is defined in Eq.~(\ref{F0}) and $\left(  \Delta
l/l\right)  _{\alpha}$ are corresponding relative changes of free domain sizes
in principal directions $\alpha=1,2,3$, see Fig.~\ref{fig1} a. The strain in
local coordinate system $\hat{\varepsilon}^{loc}\left(  {\mathbf{x}}\right)  $
is related to the strain $\hat{\varepsilon}$ in global coordinate system via
Eqs.~(\ref{eRR}) and~(\ref{matrix}). Notice, that the above defined strain
tensors have different origination: $\hat{\varepsilon}^{F}$ (Eq.~\ref{e_free})
describes irreversible irradiation induced domain dilatation and creep,
$\hat{\varepsilon}$ is truly elastic deformation tensor and the macroscopic
deformation $\hat{\varepsilon}^{M}$ (Eq.~\ref{eir}) has both inelastic and
elastic contributions.

Symmetric stiffness matrix $\hat{c}$ in Eq.~(\ref{E1}) is nearly the same for
all domains, but the axis of its symmetry randomly rotates with respect to
direction of bulk graphite orthotropy axis. Components $c_{ij}$ of local
stiffness matrix $\hat{c}$ are elastic moduli of the domain. They change under
irradiation and depend on neutron fluence and temperature (see
section~\ref{ESTIMATION} below). Here we consider the case of transversely
isotropic symmetry, when the matrix $\hat{c}$ in local coordinate system has
the form
\begin{equation}
\hat{c}=\left(
\begin{array}
[c]{cccccc}%
c_{11} & c_{12} & c_{13} & 0 & 0 & 0\\
c_{12} & c_{11} & c_{13} & 0 & 0 & 0\\
c_{13} & c_{13} & c_{33} & 0 & 0 & 0\\
0 & 0 & 0 & c_{44} & 0 & 0\\
0 & 0 & 0 & 0 & c_{44} & 0\\
0 & 0 & 0 & 0 & 0 & \left(  \allowbreak c_{11}-c_{12}\right)  /2
\end{array}
\right)  \label{c}%
\end{equation}

Substituting the local strain $\hat{\varepsilon}^{loc}$ from Eq.~(\ref{matrix}%
) to elastic energy~(\ref{E1}) we finally get elastic energy of diffuse domain
structure
\begin{equation}
E=\dfrac{1}{2}%
{\displaystyle\int}
d{\mathbf{x}%
{\displaystyle\sum\nolimits_{ij}}
}c_{ij}\left[  \hat{R}^{-1}\left(  \hat{\varepsilon}+\hat{\varepsilon}%
^{M}\right)  -\hat{\varepsilon}^{F}\right]  _{i}\left[  \hat{R}^{-1}\left(
\hat{\varepsilon}+\hat{\varepsilon}^{M}\right)  -\hat{\varepsilon}^{F}\right]
_{j} \label{E}%
\end{equation}
Using this elastic energy we describe below in section~\ref{FREE} three the
most important cases of domain structures:

\begin{description}
\item[Regular:] Regular structure (see Fig.~\ref{fig1} a) deforms affinely
with domain deformation. In this case the equilibrium macroscopic deformation
is the same as for free domains, Eq.~(\ref{e_free}), $\hat{\varepsilon}%
^{M}=\hat{\varepsilon}^{loc}=\hat{\varepsilon}^{F}$.

\item[Free:] In the case of \textquotedblleft unconnected\textquotedblright%
\ freely dilated domains $\hat{\varepsilon}^{loc}=\hat{\varepsilon}^{F}$ and
the macroscopic deformation $\hat{\varepsilon}^{M}=\overline{\hat{R}}%
\hat{\varepsilon}^{F}$ differs from deformation of individual domains
$\hat{\varepsilon}^{F}$ only due to difference in domain orientations. Here
the bar means averaging over domain orientation.

\item[Irregular:] Domains of irregular structure (see Fig.~\ref{fig1} b) are
in more cramped conditions with respect to the case of regular domain
structure. Elastic environment tends to compress the domain with respect to
the case of freely dilated domains and $\hat{\varepsilon}^{loc}\neq
\hat{\varepsilon}^{F}$. Equilibrium deformation $\hat{\varepsilon}^{M}$ is
determined by balance between elastic deformation of the domain and elastic
response of its environment on this deformation.

\item[Uniform:] In the case of uniform distribution of domain directions (see
Fig.~\ref{fig1} c) the macroscopic deformation $\hat{\varepsilon}^{M}$ is
isotropic even for strongly anisotropic deformation of individual domains,
$\hat{\varepsilon}^{loc}\neq\hat{\varepsilon}^{F}$.
\end{description}

\section{Macroscopic description of diffuse domain structure\label{MACRO}}

In this section we develop macroscopic theory of reactor graphite elasticity.
Although diffuse domains have different orientations, many such domains at
spatial scales large with respect to domain size form an orthotropic and
homogeneous material from a marcomechanics standpoint. The main feature of
diffuse domain structure is absence of cross-domain boundaries, near which the
breakup of stress or strain may take place as in ordinary polycristalline
solids. In the case of diffuse domain structure both stress and strain change
continuously and the domains of diffuse domain structure directly feel the
stress $\hat{\sigma}\left(  {\mathbf{x}}\right)  $ of the elastic enviroment.
Under the influence of such stress the randomly turned domains deform
non-affinely with macroscopic deformation of bulk graphite (see
Fig.~\ref{fig1} b and c) and the strain $\hat{\varepsilon}$ of such irregular
domain structure gains large short wavelength (on the order of domain size
$l$) components.

\subsection{Local stress}

Elastic energy~(\ref{E}) can be expanded in powers of the strain
$\hat{\varepsilon}$. We leave linear term in this expansion unchanged, and
transform only its quadratic term using the equality\footnote{Eq.~(\ref{quad})
is analogous to famous Hubbard-Strattonovich transformation, which is widely
used to introduce weakly fluctuating order parameter (analog of the stress
$\hat{\sigma}\left(  {\mathbf{x}}\right)  $) in systems with strongly
fluctuating dynamic variables (analoges of the strain $\hat{\varepsilon
}\left(  {\mathbf{x}}\right)  $). Examples of such order parameters are
magnetic moments in spin systems, superconducting order parameter in
electronic systems and so on.}
\begin{equation}%
\begin{array}
[c]{c}%
\dfrac{1}{2}%
{\displaystyle\int}
d{\mathbf{x}%
{\displaystyle\sum\nolimits_{ij}}
}c_{ij}\left(  \hat{R}^{-1}\hat{\varepsilon}\right)  _{i}\left(  \hat{R}%
^{-1}\hat{\varepsilon}\right)  _{j}=\\
-\dfrac{1}{2}{%
{\displaystyle\int}
}d{\mathbf{x}%
{\displaystyle\sum\nolimits_{ij}}
}S_{ij}\sigma_{i}\sigma_{j}+{%
{\displaystyle\int}
}d{\mathbf{x}%
{\displaystyle\sum\nolimits_{i}}
}\sigma_{i}\varepsilon_{i}%
\end{array}
\label{quad}%
\end{equation}
Here
\begin{equation}
\hat{S}\left(  \Omega\right)  =\hat{R}\left(  \Omega\right)  \hat{s}\hat
{R}^{T}\left(  \Omega\right)  \label{SO}%
\end{equation}
is local compliance matrix in global coordinate system and $\hat{s}=\hat
{c}^{-1}$ is local compliance matrix inverse to the local stiffness matrix
$\hat{c}$, Eq.~(\ref{c}). Let us show that if the stress $\hat{\sigma}$ is
found from extremum of the right hand side of Eq.~(\ref{quad}), the latter
will be equal to the left hand side of this equation. Substituting the
solution of extremum equation for the local stress
\begin{equation}
\sigma_{i}=\sum\nolimits_{j}S_{ij}^{-1}\left(  \Omega\right)  \varepsilon
_{j}=\sum\nolimits_{j}\left[  \hat{R}^{-T}\left(  \Omega\right)  \hat{c}%
\hat{R}^{-1}\left(  \Omega\right)  \right]  _{ij}\varepsilon_{j} \label{se}%
\end{equation}
back to the right hand side of Eq.~(\ref{quad}) we really reproduce the
expression for the left hand side of this equation.

Using Eq.~(\ref{quad}) we can transform quadratic in $\hat{\varepsilon}$ term
in Eq.~(\ref{E}) into corresponding linear term and rewrite elastic
energy~(\ref{E}) in terms of the local stress $\hat{\sigma}\left(
{\mathbf{x}}\right)  $:
\begin{align}
E\left[  \sigma\right]   &  =const-\dfrac{1}{2}\int d{\mathbf{x}}%
\sum\nolimits_{ij}S_{ij}\left[  \Omega\left(  {\mathbf{x}}\right)  \right]
\sigma_{i}\left(  {\mathbf{x}}\right)  \sigma_{j}\left(  {\mathbf{x}}\right)
+\nonumber\\
&  \int d{\mathbf{x}}%
{\displaystyle\sum\nolimits_{i}}
\left(  \sigma_{i}\left(  {\mathbf{x}}\right)  +\sigma_{i}^{M}\left[
\Omega\left(  {\mathbf{x}}\right)  \right]  -\sigma_{i}^{F}\left[
\Omega\left(  {\mathbf{x}}\right)  \right]  \right)  \varepsilon_{i}\left(
{\mathbf{x}}\right)  \label{Eel}%
\end{align}
where $\hat{\sigma}^{F}$ is stress because of irradiation-induced deformation
of the domain. In matrix notations
\begin{equation}
\sigma_{i}^{F}\left[  \Omega\left(  {\mathbf{x}}\right)  \right]  =\hat
{R}^{-T}\left(  \Omega\right)  \hat{c}\hat{\varepsilon}^{F} \label{sF}%
\end{equation}
Stress $\hat{\sigma}^{M}$ describes elastic response of medium on this
deformation:
\begin{equation}
\hat{\sigma}^{M}\left(  \Omega\right)  =\hat{R}^{-T}\left(  \Omega\right)
\hat{c}\hat{R}^{-1}\left(  \Omega\right)  \hat{\varepsilon}^{M}=\hat{S}%
^{-1}\left(  \Omega\right)  \hat{\varepsilon}^{M} \label{sL}%
\end{equation}

Minimum of elastic energy~(\ref{Eel}) gives standard equation of equilibrium
in the presence of external stress $\sigma_{i}\left(  {\mathbf{x}}\right)  $:
\begin{equation}%
{\displaystyle\sum\limits_{\beta}}
\frac{\partial}{\partial x_{\beta}}\left(  \sigma_{\alpha\beta}+\sigma
_{\alpha\beta}^{M}-\sigma_{\alpha\beta}^{F}\right)  =0 \label{min1}%
\end{equation}
To derive this equation we substituted elastic strain~(\ref{eab}) in
Eq.~(\ref{Eel}). Variation of elastic energy for given stress tensor
$\hat{\sigma}$ takes the form%
\[
\delta E=\int d{\mathbf{x}}%
{\displaystyle\sum\nolimits_{\alpha\beta}}
\left(  \sigma_{\alpha\beta}+\sigma_{\alpha\beta}^{M}-\sigma_{\alpha\beta}%
^{F}\right)  \frac{\partial\delta u_{\alpha}}{\partial x_{\beta}}%
\]
Integrating this expression by parts and equating volume term to zero, we find
Eq.~(\ref{min1}).

The solution of this linear equation can be written as a superposition of
contributions from all domains. Each of these contributions slowly decays with
the distance from the domain (as a power law at large distances). The domains
have random orientations, and their total contribution to the stress is
effectively self-averaged because of discussed above long-range character of
elasticity. Therefore, in deriving equations of macroscopic elasticity we may
consider the stress $\hat{\sigma}\left(  {\mathbf{x}}\right)  $ as weakly
varying on the scale of domain size function. Since angles $\Omega$ randomly
alter in space, the constant stress $\hat{\sigma}$ describes strain
$\hat{\varepsilon}\left(  {\mathbf{x}}\right)  $ strongly varying in space on
the scale of domain size, see Eq.~(\ref{se}).

Note that Eq.~(\ref{min1}) corresponds to minimum of energy~(\ref{Eel}).
Because of this minimum condition substitution of the approximate solution
$\hat{\sigma}\left(  {\mathbf{x}}\right)  \simeq const$ of stochastic
equation~(\ref{min1}) in energy~(\ref{Eel}) gives only minor corrections of
second order in small-scale space variations of the function $\hat{\sigma
}\left(  {\mathbf{x}}\right)  $.

\subsection{Derivation of macroscopic elasticity equations}

Below we will neglect the effect of weak spatial stress variations on the
scale of domain size $l$ on macroscopic elasticity of graphite. Than the
integration over ${\mathbf{x}}$ in Eq.~(\ref{Eel}) is equivalent to
pre-averaging of the stress expansion coefficients $S_{ij}\left[
\Omega\left(  {\mathbf{x}}\right)  \right]  ,\sigma_{i}^{M}\left[
\Omega\left(  {\mathbf{x}}\right)  \right]  $ and $\sigma_{i}^{F}\left[
\Omega\left(  {\mathbf{x}}\right)  \right]  $ over characteristic spatial
scale of the function $\hat{\sigma}\left(  {\mathbf{x}}\right)  $, and
macroscopic free energy for slowly varying stress $\hat{\sigma}\left(
{\mathbf{x}}\right)  $ takes the form
\begin{equation}%
\begin{array}
[c]{c}%
E\left[  \sigma\right]  =-\dfrac{1}{2}{%
{\displaystyle\int}
}d{\mathbf{x}%
{\displaystyle\sum\nolimits_{ij}}
}\bar{S}_{ij}\sigma_{i}\left(  {\mathbf{x}}\right)  \sigma_{j}\left(
{\mathbf{x}}\right) \\
+{%
{\displaystyle\int}
}d{\mathbf{x}%
{\displaystyle\sum\nolimits_{i}}
}\left[  \sigma_{i}\left(  {\mathbf{x}}\right)  +\bar{\sigma}_{i}^{M}\left(
{\mathbf{x}}\right)  -\bar{\sigma}_{i}^{F}\left(  {\mathbf{x}}\right)
\right]  \varepsilon_{i}\left(  {\mathbf{x}}\right)  ,
\end{array}
\label{Es}%
\end{equation}
where bar means averaging over space.

Equation of equilibrium~(\ref{min1}) can also be transferred to its
pre-averaged form with the aid of equality
\[%
\begin{array}
[c]{c}%
\overline{\dfrac{\partial F\left(  {\mathbf{x}}\right)  }{\partial x_{\beta}}%
}={%
{\displaystyle\int}
}d{\mathbf{x}}^{\prime}w\left(  {\mathbf{x}}^{\prime}-{\mathbf{x}}\right)
\dfrac{\partial F\left(  {\mathbf{x}}^{\prime}\right)  }{\partial x_{\beta
}^{\prime}}=-{%
{\displaystyle\int}
}d{\mathbf{x}}^{\prime}\dfrac{\partial w\left(  {\mathbf{x}}^{\prime
}-{\mathbf{x}}\right)  }{\partial x_{\beta}^{\prime}}F\left(  {\mathbf{x}%
}\right) \\
=\dfrac{\partial}{\partial x_{\beta}}{%
{\displaystyle\int}
}d{\mathbf{x}}^{\prime}w\left(  {\mathbf{x}}^{\prime}-{\mathbf{x}}\right)
F\left(  {\mathbf{x}}^{\prime}\right)  =\dfrac{\partial\bar{F}\left(
{\mathbf{x}}\right)  }{\partial x_{\beta}}%
\end{array}
\]
which is valid for any function $F\left(  {\mathbf{x}}\right)  $ and bell-like
weight function $w\left(  {\mathbf{x}}^{\prime}-{\mathbf{x}}\right)  $. The
weight function is centered at point ${\mathbf{x}}$ and has characteristic
size larger than $l$. As the result of such averaging Eq.~(\ref{min1}) for
slowly varying stress $\hat{\sigma}\left(  {\mathbf{x}}\right)  $ takes the
form
\begin{equation}
\sum\limits_{\beta}\frac{\partial}{\partial x_{\beta}}\left[  \sigma
_{\alpha\beta}\left(  {\mathbf{x}}\right)  +\bar{\sigma}_{\alpha\beta}%
^{M}\left(  {\mathbf{x}}\right)  -\bar{\sigma}_{\alpha\beta}^{F}\left(
{\mathbf{x}}\right)  \right]  =0 \label{ss}%
\end{equation}
In equilibrium neighbouring domains can self-consistently adapt their
deformations since rigid directions of these domains are deployed by random
angles to each other. Eq.~\ref{ss} should be solved with the boundary
conditions on surface of bulk sample:%
\begin{equation}
\sum\nolimits_{\beta}\left[  \sigma_{\alpha\beta}\left(  {\mathbf{x}}\right)
+\bar{\sigma}_{\alpha\beta}^{M}\left(  {\mathbf{x}}\right)  \right]  n_{\beta
}=P_{\alpha}%
\end{equation}
where $n_{\beta}$ are components of normal to the surface, and $P_{\alpha}$
are components of external force applied to the surface. This force can be
considered as a substitute at the surface of corresponding volume source of
deformation $\bar{\sigma}_{\alpha\beta}^{F}$ in Eq.~\ref{ss}.

Elastic energy~(\ref{Es}) can be rewritten in standard form of macroscopic
theory of elasticity by introducing \textquotedblleft
smoothed\textquotedblright\ on the domain scale $l$ strain tensor
$\bar{\varepsilon}$ with components
\begin{equation}
\bar{\varepsilon}_{i}\left(  {\mathbf{x}}\right)  =\sum\nolimits_{j}\bar
{S}_{ij}\sigma_{j}\left(  {\mathbf{x}}\right)  \label{eSs}%
\end{equation}
which is obtained by averaging the microscopic strain $\hat{\varepsilon}$ with
the above weight function $w$:
\[%
\begin{array}
[c]{c}%
\bar{\varepsilon}_{i}\left(  {\mathbf{x}}\right)  =%
{\displaystyle\int}
d{\mathbf{x}}^{\prime}w\left(  {\mathbf{x}}^{\prime}-{\mathbf{x}}\right)
\varepsilon_{i}\left(  {\mathbf{x}}^{\prime}\right) \\
=%
{\displaystyle\int}
d{\mathbf{x}}^{\prime}w\left(  {\mathbf{x}}^{\prime}-{\mathbf{x}}\right)  {%
{\displaystyle\sum\nolimits_{j}}
}S_{ij}\left[  \Omega\left(  {\mathbf{x}}^{\prime}\right)  \right]  \sigma
_{j}\left(  {\mathbf{x}}^{\prime}\right) \\
\simeq%
{\displaystyle\int}
d{\mathbf{x}}^{\prime}w\left(  {\mathbf{x}}^{\prime}-{\mathbf{x}}\right)  {%
{\displaystyle\sum\nolimits_{j}}
}S_{ij}\left[  \Omega\left(  {\mathbf{x}}^{\prime}\right)  \right]  \sigma
_{j}\left(  {\mathbf{x}}\right)  =\sum_{j}\bar{S}_{ij}\sigma_{j}\left(
{\mathbf{x}}\right)
\end{array}
\]
Substituting Eq.~(\ref{eSs}) into elastic energy~(\ref{Es}) we can rewrite it
in standard form for macroscopic elasticity:
\begin{align}
E\left[  \bar{\varepsilon}\right]   &  =%
{\displaystyle\int}
d{\mathbf{x}}\left[  \dfrac{1}{2}{%
{\displaystyle\sum\nolimits_{ij}}
}\bar{C}_{ij}\bar{\varepsilon}_{i}\left(  {\mathbf{x}}\right)  \bar
{\varepsilon}_{j}\left(  {\mathbf{x}}\right)  +\right. \nonumber\\
&  \left.  {%
{\displaystyle\sum\nolimits_{i}}
}\left(  \bar{\sigma}_{i}^{M}\left(  {\mathbf{x}}\right)  -\bar{\sigma}%
_{i}^{F}\left(  {\mathbf{x}}\right)  \right)  \bar{\varepsilon}_{i}\left(
{\mathbf{x}}\right)  \right]  \label{Em}%
\end{align}
where $\bar{C}=\bar{S}^{-1}$ is macroscopic stiffness matrix, $\bar{\sigma
}_{i}^{F}\left(  {\mathbf{x}}\right)  $ and $\bar{\sigma}_{i}^{M}\left(
{\mathbf{x}}\right)  $ are averages of the corresponding values~(\ref{sF})
and~(\ref{sL}). Minimum of this energy reproduces equation of equilibrium of
effective elastic medium~(\ref{ss}), where the stress $\hat{\sigma}$ is
related to the strain $\bar{\varepsilon}$ via Eq.~(\ref{eSs}). Since we
defined $\hat{\varepsilon}^{M}$ as equilibrium macroscopic strain (see
Eq.~\ref{eir}) in equilibrium we have $\sigma_{\alpha\beta}\left(
{\mathbf{x}}\right)  =\bar{\varepsilon}_{\alpha\beta}\left(  {\mathbf{x}%
}\right)  =0$. Eq.~(\ref{Em}) with $\bar{\varepsilon}_{\alpha\beta}\left(
{\mathbf{x}}\right)  \neq0$ describing elastic response of graphite to
external perturbation (elastic moduli, see next section).

\subsection{Elastic moduli of bulk sample\label{MODULI}}

The average compliance matrix $\bar{S}$ in Eq.~(\ref{Es}) is found by space
averaging the local compliance matrix $\hat{S}\left(  \Omega\right)  $,
Eq.~(\ref{SO}):
\begin{equation}
\bar{S}_{ij}=\sum\nolimits_{kl}\overline{R_{ik}\left(  \Omega\right)
R_{jl}\left(  \Omega\right)  }s_{kl} \label{sav}%
\end{equation}
Averaging of inverse stiffness matrix, and not of stiffness matrix itself, is
related to the fact that domains deform non-affinely with the macroscopic
strain, see Fig.~\ref{fig1} b\cite{Voigt-89}. Instead, they directly feel the
applied macroscopic stress $\hat{\sigma}\left(  {\mathbf{x}}\right)
$\cite{Reuss-29}.

Space averaging in Eqs.~(\ref{Es}) and~(\ref{sav}) can be rewritten as
averaging over all orientations of domains $\Omega=\left(  \varphi,\theta
,\psi\right)  $. Rotation angles $\varphi$ and $\psi$ are randomly distributed
in the interval $\left(  0,2\pi\right)  $ while probability distribution of
the azimuthal angle $\theta$ can be characterized by the second and fourth
moments of $\sin\theta$:
\begin{equation}
m_{2}=\int_{0}^{\pi/2}\sin^{2}\theta\Psi\left(  \theta\right)  \sin\theta
d\theta,\quad m_{4}=\int_{0}^{\pi/2}\sin^{4}\theta\Psi\left(  \theta\right)
\sin\theta d\theta\label{m2m4}%
\end{equation}
The distribution function $\Psi\left(  \theta\right)  $ of the angle $\theta$
is calculated below in section~\ref{DISTRIBUTION} in case of graphite
manufactured through the process of extrusion. In this case the fourth moment
$m_{4}$ can be expressed in terms of the second moment $m_{2}$, see
Eq.~\ref{m24x} below.

Calculating the angular average in Eq.~(\ref{sav}) we get linear relation
between elements of averaged and local compliance matrixes:
\begin{align}
\bar{S}_{11}  &  =s_{11}+\tfrac{1}{2}\left(  -2s_{11}+2s_{13}+s_{44}\right)
m_{2}+\tfrac{3}{8}s,\nonumber\\
\bar{S}_{12}  &  =s_{12}+\left(  -s_{12}+s_{13}\right)  m_{2}+\tfrac{1}%
{8}s,\nonumber\\
\bar{S}_{13}  &  =s_{13}+\tfrac{1}{2}\left(  s_{11}+s_{12}-3s_{13}%
+s_{33}-s_{44}\right)  m_{2}-\tfrac{1}{2}s,\label{s1}\\
\bar{S}_{33}  &  =s_{33}+\left(  2s_{13}-2s_{33}+s_{44}\right)  m_{2}%
+s,\nonumber\\
\bar{S}_{44}  &  =s_{44}+\tfrac{1}{2}\left(  6s_{11}-2s_{12}-8s_{13}%
+4s_{33}-5s_{44}\right)  m_{2}-2s,\nonumber\\
s  &  =\left(  s_{11}-2s_{13}+s_{33}-s_{44}\right)  \allowbreak m_{4}\nonumber
\end{align}

In case of uniform distribution of domain directions $\Psi\left(
\theta\right)  =1$ substituting the moments
\begin{equation}
m_{2}=2/3,\qquad m_{4}=8/15 \label{m_unif}%
\end{equation}
into the above equations we obtain moduli of macroscopically isotropic
material (see Fig.~\ref{fig1} c):
\begin{align}
\bar{S}_{11}  &  =\bar{S}_{33}=\tfrac{1}{15}\left(  8s_{11}+4s_{13}%
+3s_{33}+2s_{44}\right)  ,\nonumber\\
\bar{S}_{12}  &  =\bar{S}_{13}=\tfrac{1}{15}\left(  s_{11}+5s_{12}%
+8s_{13}+s_{33}-s_{44}\right)  ,\label{unif}\\
\bar{S}_{44}  &  =2\left(  \bar{S}_{11}-\bar{S}_{12}\right)  =\tfrac{2}%
{15}\left(  7s_{11}-5s_{12}-4s_{13}+2s_{33}+5s_{44}\right) \nonumber
\end{align}

Anisotropy and degree of disorientation increase with decreasing spatial scale
and become large on scale of a stack of microcrystallites. Under the effect of
thermal expansion of microcrystallites amplitude of stress on this scale can
reach the level of crack initiation and growth. The characteristic scale of
emerging microcracks is small compared to domain size. Therefore,
relation~(\ref{s1}) between macroscopic and domain modules do not depend on
evolution of microcrack subsystem and is determined only by the distribution
of domain orientations.

\subsection{Free boundary conditions\label{FREE}}

Solution of Eq.~(\ref{ss}) for samples subjected to external pressure $P$ is
$\bar{\sigma}^{M}-\bar{\sigma}^{F}=\hat{\sigma}^{ext}$ with diagonal tensor
$\sigma_{\alpha\beta}^{ext}=P\delta_{\alpha\beta}$. In case of free boundary
conditions there is no stress at the boundary of the sample, $\hat{\sigma
}^{ext}=0$, and the equilibrium macroscopic strain $\hat{\varepsilon}^{M}$ is
determined by the condition of zero average stress:
\begin{equation}
\bar{\sigma}^{M}=\bar{\sigma}^{F} \label{str0}%
\end{equation}
This condition corresponds to equilibrium between the average stress due to
irradiation-induced deformation of domains, Eq.~(\ref{sF}) and the average
stress due to response of elastic medium on this deformation, Eq.~(\ref{sF}).
Averaging the local stresses, Eqs.~(\ref{sF}) and~(\ref{sL}), over Euler
angles $\Omega=\left(  \psi,\phi,\theta\right)  $ we find linear matrix
equation for the macroscopic strain~(\ref{eir}):
\[
\overline{\hat{S}^{-1}\left(  \Omega\right)  }\hat{\varepsilon}^{M}%
=\overline{\hat{R}^{-T}\left(  \Omega\right)  }\hat{c}\hat{\varepsilon}^{F}%
\]
The solution of this equation is%
\begin{align}
\left(  \Delta L/L\right)  ^{\parallel}  &  =\left(  \Delta l/l\right)
_{\parallel}+k^{\parallel}\left[  \left(  \Delta l/l\right)  _{\perp}-\left(
\Delta l/l\right)  _{\parallel}\right]  ,\nonumber\\
\ \left(  \Delta L/L\right)  ^{\perp}  &  =\left(  \Delta l/l\right)  _{\perp
}-\tfrac{1}{2}k^{\perp}\left[  \left(  \Delta l/l\right)  _{\perp}-\left(
\Delta l/l\right)  _{\parallel}\right]  \label{eav}%
\end{align}
The factors $k$ in this expression can be written in explicit form:
\begin{align}
k^{\parallel}  &  =\left[  2\left(  4c_{44}c_{\parallel}-\Delta\right)
\allowbreak m_{2}+\left(  c_{\perp}-c_{\parallel}\right)  \allowbreak\gamma
m_{2}^{2}\right. \nonumber\\
&  \left.  \left.  +2c_{\parallel}\left(  \gamma-4c_{44}\right)  \allowbreak
m_{4}\right]  \right/  K,\label{k1}\\
k^{\perp}  &  =\left[  2\left(  4c_{44}c_{\perp}-\Delta\right)  m_{2}+2\left(
c_{\parallel}-c_{\perp}\right)  \left(  \gamma+c_{\perp}-c_{\parallel}\right)
\allowbreak m_{2}^{2}\right. \nonumber\\
&  \left.  \left.  +2c_{\perp}\left(  \gamma-4c_{44}\right)  \allowbreak
m_{4}\right]  \right/  K,\label{k2}\\
K  &  =2\Delta+\left[  4c_{44}\left(  c_{\perp}+2c_{\parallel}\right)
-6\Delta\right]  \allowbreak m_{2}-\left(  c_{\perp}-c_{\parallel}\right)
^{2}m_{2}^{2}\nonumber\\
&  +\allowbreak\left(  c_{\perp}+2c_{\parallel}\right)  \left(  \gamma
-4c_{44}\right)  \allowbreak\allowbreak m_{4}\nonumber
\end{align}
where $m_{2}$ and $m_{4}$ are moments of $\sin\theta$ (see Eq.~\ref{m2m4}) and
we use abbreviations
\[%
\begin{array}
[c]{ll}%
c_{\parallel}=c_{11}+c_{12}+c_{13}, & c_{\perp}=2c_{13}+c_{33},\\
\gamma=c_{11}+c_{12}-2c_{13}, & \Delta=c_{33}\left(  c_{11}+c_{12}\right)
-2c_{13}^{2}%
\end{array}
\]

Combining Eqs.~(\ref{eav}) we find radiation-induced change of sample volume
\begin{align}
\Delta V/V  &  =\left(  \Delta L/L\right)  ^{\parallel}+2\left(  \Delta
L/L\right)  ^{\perp}\nonumber\\
=  &  \left(  \Delta l/l\right)  _{\parallel}+2\left(  \Delta l/l\right)
_{\perp}-k_{V}\left[  \left(  \Delta l/l\right)  _{\perp}-\left(  \Delta
l/l\right)  _{\parallel}\right]  , \label{dV}%
\end{align}
Amplitude of volume change is described by the factor $k_{V}$:
\begin{align}
k_{V}  &  =k^{\perp}-k^{\parallel}=\left(  c_{\perp}-c_{\parallel}\right)
\left[  8c_{44}m_{2}-\left(  3\gamma+2c_{\perp}-2c_{\parallel}\right)
m_{2}^{2}\right. \nonumber\\
&  \left.  \left.  +2\left(  \gamma-4c_{44}\right)  m_{4}\right]  \right/  K
\label{kV}%
\end{align}

In case $k_{V}=0$ ($c_{\perp}=c_{\parallel}$) the relative volume change is
the same as for medium with freely dilating domains
\[
\left(  \Delta V/V\right)  _{F}=\left(  \Delta l/l\right)  _{\parallel
}+2\left(  \Delta l/l\right)  _{\perp}%
\]
although such sample deforms strongly anisotropically (see Fig.~\ref{fig1} b).
In case $k_{V}\neq0$ the difference between those volume changes is related to
deformation of domain shape due to elastic response of environment. The case
$k_{V}>0$ corresponds to compression of the domain by the medium, and
$k_{V}<0$ describes its dilation.

Consider most important cases in more detail:

\begin{description}
\item[Regular:] Regular domain structure corresponds to case $m_{2}=m_{4}=0$,
when the macroscopic strain coincides with the irradiation-induced strain,
\[
\left(  \Delta L/L\right)  ^{\parallel}=\left(  \Delta l/l\right)
_{\parallel},\qquad\left(  \Delta L/L\right)  ^{\perp}=\left(  \Delta
l/l\right)  _{\perp}%
\]
The bulk sample deforms affinely with domain deformation, see Fig.~\ref{fig1} a.

\item[Free:] Stress in freely dilated domains is determined by the local
condition $\hat{\sigma}^{M}\left(  {\mathbf{x}}\right)  =\hat{\sigma}%
^{F}\left(  {\mathbf{x}}\right)  $ that does not satisfy the equilibrium
equation~(\ref{min1}) due to long-range character of elasticity. The solution
$\hat{\varepsilon}^{M}\left(  {\mathbf{x}}\right)  =\hat{R}\left[
\Omega\left(  {\mathbf{x}}\right)  \right]  \hat{\varepsilon}^{F}$ of the
above condition takes into account only domain rotation and ignores elastic
response of the medium on irradiation-induces deformation of domains. Below
for comparison we write down macroscopic deformation of such fictional medium
of freely dilating objects:%
\begin{align}
\ \left(  \Delta L/L\right)  _{F}^{\parallel}  &  =\left(  \Delta l/l\right)
_{\parallel}+m_{2}\left[  \left(  \Delta l/l\right)  _{\perp}-\left(  \Delta
l/l\right)  _{\parallel}\right] \nonumber\\
\left(  \Delta L/L\right)  _{F}^{\perp}  &  =\left(  \Delta l/l\right)
_{\perp}-\tfrac{1}{2}m_{2}\left[  \left(  \Delta l/l\right)  _{\perp}-\left(
\Delta l/l\right)  _{\parallel}\right]  \label{free}%
\end{align}
and corresponding volume change of such medium:
\begin{equation}
\left(  \frac{\Delta V}{V}\right)  _{F}=\left(  \frac{\Delta L}{L}\right)
_{F}^{\parallel}+2\left(  \frac{\Delta L}{L}\right)  _{F}^{\perp}=\left(
\frac{\Delta l}{l}\right)  _{\parallel}+2\left(  \frac{\Delta l}{l}\right)
_{\perp} \label{dVF}%
\end{equation}

\item[Irregular:] At high anisotropy of domain directions (see Fig.~\ref{fig1}
b) the factors $k$ are proportional to the second moment $m_{2}\ll1$:
\begin{equation}
k^{\parallel}\simeq\left(  4c_{44}c_{\parallel}/\Delta-1\right)  m_{2},\quad
k^{\perp}\simeq\left(  4c_{44}c_{\perp}/\Delta-1\right)  m_{2} \label{ksmall}%
\end{equation}
Nonaffine deformation of bulk sample is related both to random rotation of
domains and elastic response of the medium on their shape-change. To separate
both contributions compare Eqs.~(\ref{eav}) and~(\ref{ksmall}) with
Eqs.~(\ref{free}). The factors $k^{\parallel}$ and $k^{\perp}$ are both
proportional to $m_{2}$, and similarly to Eq.~(\ref{free}) they take into
account the effect of domain disorientation on sample size. The difference
between $k^{\parallel},k^{\perp}$ and $m_{2}$ is due to compression of domains
by their elastic environment.

\item[Uniform:] In case of uniform distribution of domain directions (see
Fig.~\ref{fig1} c) with moments~(\ref{m_unif}) we find from Eqs.~(\ref{k1})
and~(\ref{k2}) that the sample as the whole deforms isotropically under
irradiation:
\[%
\begin{array}
[c]{c}%
\left(  \Delta L/L\right)  ^{\parallel}=\left(  \Delta L/L\right)  ^{\perp
}=\ \\
\dfrac{c_{\perp}}{c_{\perp}+2c_{\parallel}}\left(  \dfrac{\Delta l}{l}\right)
_{\parallel}+\dfrac{2c_{\parallel}}{c_{\perp}+2c_{\parallel}}\left(
\dfrac{\Delta l}{l}\right)  _{\perp}\
\end{array}
\]

\end{description}

\section{Formation of diffuse domain structure\label{FORMATION}}

In this section we propose simple model of diffuse domains formation in
graphite prepared by direct melt extrusion process. We also estimate the size
of such domains and describe orientational alignment of domains in extrusion flow.

\subsection{Typical domain size $l$}

We will treat a mixture of pitch and coke particles at fabrication conditions
as very viscous incompressible fluid. Velocity field ${\mathbf{v}}%
({\mathbf{x}})$ of such fluid with density $\rho$ can be described by
Navier-Stokes equation:
\begin{equation}
\rho\left[  \partial{\mathbf{v}}/\partial t+\left(  {\mathbf{v}}%
\triangledown\right)  {\mathbf{v}}\right]  -\eta\triangledown^{2}%
{\mathbf{v=f}} \label{NS}%
\end{equation}
with boundary condition ${\mathbf{v}}=0$ at channel walls. Here $\triangledown
$ is gradient, $\eta$ is dynamic viscosity of the fluid and force
${\mathbf{f}}$ initiates rotation of random graphite crystallites. Before
extrusion the crystallites have random directions of anisotropy (with uniform
distribution function $\Psi_{0}\left(  \theta\right)  =1$), and they are
rotated in the flow under influence of gradient of the force ${\mathbf{f}}$.
Mechanism of such rotation is different on spatial scales small and large with
respect to characteristic correlation length
\begin{equation}
l{\mathbf{\simeq\operatorname{Re}}}\eta/\left(  \rho v\right)  \label{l0}%
\end{equation}
where $\operatorname{Re}$ is effective Reynolds number:

\begin{description}
\item[At small scales $x<l$] main contribution to left hand side of
Eq.~(\ref{NS}) comes from viscous term $-\eta\triangledown^{2}{\mathbf{v\simeq
f}}$. This term stabilizes the flow inside the correlation volume $l^{3}$ and
describes regular laminar motion of the fluid on the scale $l$. Therefore, the
rotation of anisotropy axes due to the force ${\mathbf{f}}$ is strongly
correlated at small spatial scales $x<l$.

\item[At large scales $x>l$] main contribution to the turning force
${\mathbf{f}}$ comes from the term $\rho\left(  {\mathbf{v}}\triangledown
\right)  {\mathbf{v\simeq f}}$, which describes convective acceleration
(instability) of the flow and leads to randomization of motion at large
distances $x>l$ due to amplification of the effect of random initial
orientation of crystallites. This amplification also leads to a decrease in
the Reynolds number $\operatorname{Re}$ with respect to case of ordinary
structureless fluid.
\end{description}

Using characteristic Reynolds number $\operatorname{Re}\simeq10$ we find from
Eq.~(\ref{l0}) a simple estimate of domain size for viscous fluid with typical
kinematic viscosity $\nu=\eta/\rho\simeq100\operatorname{cSt}$
($\operatorname{mm}^{2}/\operatorname{s}$) and average velocity of flow
$v\simeq10\operatorname{cm}/\operatorname{s}$:
\begin{equation}
l\simeq\operatorname{Re}\nu/v\simeq1\operatorname{cm} \label{lRe}%
\end{equation}
This estimation is in good agreement with domain size $l\simeq
0.6\operatorname{cm}$ observed in GR-280 graphite\cite{16n}. Although our
estimate of the Reynolds number $\operatorname{Re}$ needs additional
experimental justification, Eq.~(\ref{lRe}) can be used to predict flow
temperature and velocity gradient dependence of the domain size $l$ in reactor graphites.

\subsection{Distribution of domain orientations\label{DISTRIBUTION}}

\begin{figure}
[tbh]
\begin{center}
\includegraphics[
height=2.165in,
width=3.0285in
]%
{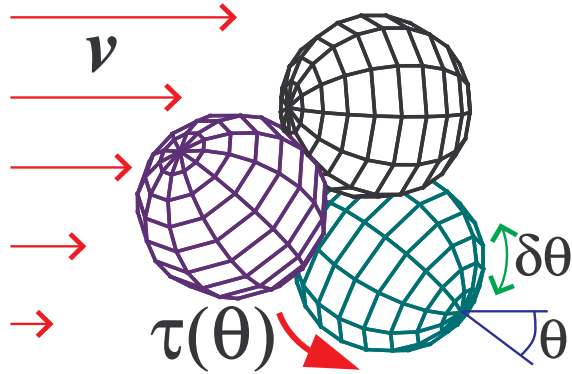}%
\caption{Randomly turned coarse-grained at scale $l$ domains rotate in the
gradient flow $\mathbf{v}\left(  \mathbf{x}\right)  $ at random angles
$\delta\theta$ during a small time interval $\delta t$ and are oriented under
the influence of orientation-dependent torque $\tau\left(  \theta\right)  $.}%
\label{tau1}%
\end{center}
\end{figure}

Due to random shape and orientations of crystallites the turning force
${\mathbf{f}}$ in Eq.~(\ref{NS}) is also random, and coarse-grained at scale
$l$ domains in the flow experience rotations at random angles $\delta\theta$
during a small time interval $\delta t$, see Fig.~\ref{tau1}.%
Such diffusion-like rotations can be described by orientational Smoluchowski
equation for distribution function $\Psi\left(  \theta\right)  $ of domain
angles $\theta$:
\begin{equation}
\frac{\partial\Psi\left(  \theta\right)  }{\partial t}=\frac{1}{\sin\theta
}\frac{\partial}{\partial\theta}\left\{  D\left(  \theta\right)  \sin
\theta\left[  \frac{\partial\Psi\left(  \theta\right)  }{\partial\theta}%
+\tau\left(  \theta\right)  \Psi\left(  \theta\right)  \right]  \right\}
\label{SM}%
\end{equation}
with uniform initial distribution $\Psi_{0}\left(  \theta\right)  =1$ before
extrusion. In general, the angular diffusion coefficient $D\left(
\theta\right)  =\left\langle \delta\theta^{2}\right\rangle /(2\delta t)$ can
depend on the angle $\theta$. The drift term in this equation reflects the
fact that the domain experiences an average torque $\tau\left(  \theta\right)
$ due to the flow, which restricts its free rotation. Both opposite directions
of domain anisotropy are equivalent, $\tau\left(  \pi-\theta\right)
=\tau\left(  \theta\right)  $ and therefore, the expansion of this function in
a Fourier series has the form:
\[
\tau\left(  \theta\right)  =\sum\nolimits_{n>0}\tau_{n}\sin\left(
2n\theta\right)  .
\]
Higher order corrections $\tau_{n}$ with $n>1$ are not important for our
consideration and will be dropped below. Amplitude $\tau=\tau_{1}$ of the
dimensionless torque is proportional to average strain rate of the flow.

Steady state solution of Eq.~(\ref{SM}) at times $t\gg D^{-1}$ has the form
\begin{equation}
\Psi\left(  \theta\right)  =e^{\tau\cos^{2}\theta-g\left(  \tau\right)
},\qquad g\left(  \tau\right)  =\ln\left(  \int_{0}^{\pi/2}e^{\tau\cos
^{2}\theta}\sin\theta d\theta\right)  \label{distr}%
\end{equation}
Substituting the distribution function~(\ref{distr}) in Eq.~(\ref{m2m4}) we
find expressions for first two moments of this distribution:
\begin{equation}
m_{2}=1-g^{\prime}\left(  \tau\right)  ,\qquad m_{4}=m_{2}^{2}+g^{\prime
\prime}\left(  \tau\right)  \label{m24}%
\end{equation}
In case of graphite obtained by extrusion at large velocity gradients the
torque $\tau$ is large, $\tau\gg1$, and we get from Eq.~(\ref{m24})
$m_{2}\simeq1/\tau\ll1$ and $m_{4}\simeq2m_{2}^{2}$. In case of graphite
formed under hydrostatic pressure average torque is small, $\tau\ll1$, and we
reproduce moments of uniform distribution, Eq.~(\ref{m_unif}). A typical value
of the second moment of graphite GR-280 $m_{2}\simeq0.3$ corresponds to
intermediate case $\tau\simeq4$ between those two limits.

In the Ref.\cite{16n} we show that the second moment $m_{2}$ can be directly
obtained from analysis of experiments measuring irradiation-induced
deformation of small size graphite samples. The fourth moment $m_{4}$ can be
expressed in terms of the second moment $m_{2}$ excluding parameter $\tau$
from Eqs.~(\ref{m24}). Implicit dependence~(\ref{m24}) of the fourth moment on
the second moment can be interpolated for the whole range of torque $\tau>0$
by the following simple expression:
\begin{equation}
m_{4}\simeq\frac{\allowbreak2m_{2}^{2}}{1+3m_{2}^{2}/2} \label{m24x}%
\end{equation}

\section{Dose dependence\label{ESTIMATION}}

Compliance moduli of a domain $s_{ij}$ can be estimated using experimentally
measured macroscopic compliance moduli $\bar{S}_{ij}$. Difficulties in
determining a set of five independent constants $\bar{S}_{ij}$ characterizing
elastic properties of transversely isotropic solid lie in the fact that, as a
rule, only two Young's modules $E_{\parallel},E_{\perp}$ and Poisson's ratio
$\nu$ are investigated in studies of irradiation effects of reactor graphite.
In our case, we need to know all the five independent constants as well as
their dose dependence.

Experimental data for non-irradiated Pile Grade A reactor graphite\cite{n+1}
are shown in Table~\ref{T1}. \begin{table}[th]
\centering%
\begin{tabular}
[c]{c||ccccc}%
Pile & $\bar{S}_{11}$ & $\bar{S}_{12}$ & $\bar{S}_{13}$ & $\bar{S}_{33}$ &
$\bar{S}_{44}$\\
Grade A & $2.150$ & $-0.127$ & $-0.123$ & $1.087$ & $3.333$\\\hline
\ Mono- & $s_{11}^{m}$ & $s_{12}^{m}$ & $s_{13}^{m}$ & $s_{33}^{m}$ &
$s_{44}^{m}$\\
crystal & $0.011$ & $-0.00045$ & $-0.025$ & $0.326$ & $4.35$\\\hline
\ Domain & $s_{11}$ & $s_{12}$ & $s_{13}$ & $s_{33}$ & $s_{44}$\\
& $2.\,\allowbreak5$ & $-0.12$ & $-0.19$ & $0.64$ & $2.73$%
\end{tabular}
\caption{Compliance moduli of Pile Grade A \cite{n+1} and monocrystal
non-irradiated graphites (in units of $10^{-11}\operatorname{cm^{2}/dyn}$). In
last raw we show our estimate for domain complience moduli.}%
\label{T1}%
\end{table}We expect that this graphite forms macroscopically homogeneous
domain structure similarly to GR-280 graphite, since both of them are
fabricated by the same extrusion method (see section~\ref{FORMATION} above).
Values of moments $m_{2}$ and $m_{4}$ and domain size $l$ for these graphites
may differ.

Effect of irradiation on elasticity of Pile Grade A graphite has been studied
in details only for two Young's moduli $E_{\parallel},E_{\perp}$ and Poisson's
ratios $\nu_{\parallel}$ and $\nu_{\perp}$ which are related to the
macroscopic compliance moduli as\cite{18n,19n}:%
\[%
\begin{array}
[c]{llllll}%
E_{\parallel} & \!\!\!\!=1/\bar{S}_{33}, & G_{\parallel} & \!\!\!\!=1/\bar
{S}_{44}, & \nu_{\parallel} & \!\!\!\!=-\bar{S}_{13}/\bar{S}_{33},\\
E_{\perp} & \!\!\!\!=1/\bar{S}_{11}, & G_{\perp} & \!\!\!\!=1\left/  \left[
2\left(  \bar{S}_{11}-\bar{S}_{12}\right)  \right]  \right.  , & \nu_{\perp} &
\!\!\!\!=-\bar{S}_{12}/\bar{S}_{11}.
\end{array}
\]
Dose dependence of entire set of compliance moduli has been established only
for monocrystalline and pyrolytic graphites over a wide temperature
range\cite{n+4,20n,n+6}. It is noticed that only \textquotedblleft
small\textquotedblright\ elastic constants $c_{33}$ and $c_{44}$ experience
significant changes (increase in value) with irradiation dose.

\begin{figure}
[tbh]
\begin{center}
\includegraphics[
height=3.8624in,
width=4.0515in
]%
{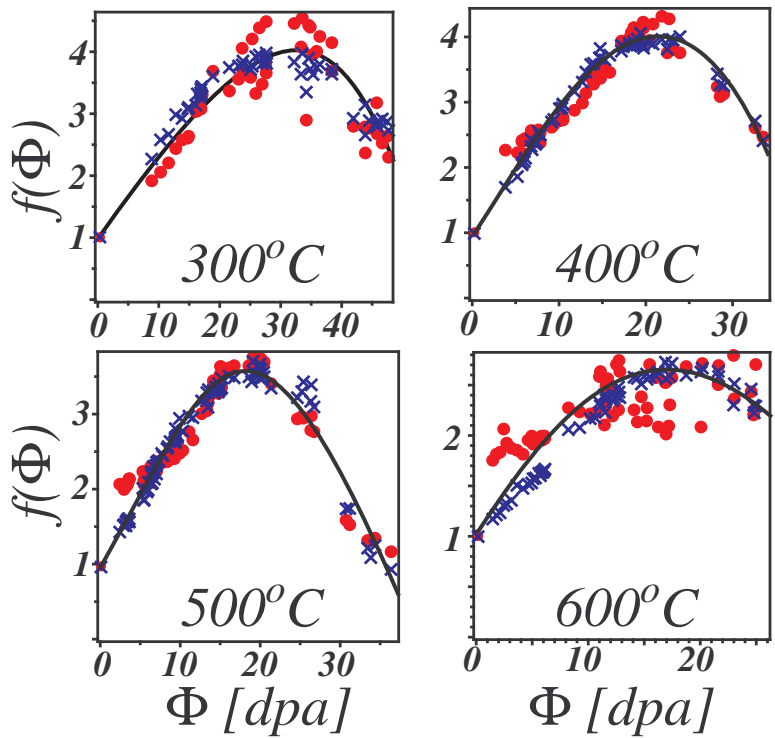}%
\caption{Dose $\Phi$ (in units of displacements per atom) dependence $f\left(
\Phi\right)  =E\left(  \Phi\right)  /E\left(  0\right)  $ of elastic
moduli\cite{n+7} of ATR-2E graphite at irradiation temperatures $300^{\circ
},400^{\circ},500^{\circ}$ and $600^{\circ}C$ (circles). Crosses show the
approximation of the function $f\left(  \Phi\right)  $ by Eq.~(\ref{ff}).}%
\label{Doses}%
\end{center}
\end{figure}

Detailed data processing for the elastic moduli $E_{\parallel}$ and $E_{\perp
}$ of ATR-2E graphite produced by extrusion, is given in Ref. \cite{n+7}. It
is shown that dose dependence of both Young's moduli $E_{\parallel}$ and
$E_{\perp}$ is the same. The only difference is in normalization factors:
\begin{equation}
E_{\parallel}\left(  \Phi\right)  =E_{\parallel}f\left(  \Phi\right)  ,\qquad
E_{\perp}\left(  \Phi\right)  =E_{\perp}f\left(  \Phi\right)  \label{E2}%
\end{equation}
Functional dependence of structure factor $f\left(  \Phi\right)  $ on neutron
fluence $\Phi$ is shown in Fig.~\ref{Doses} for different irradiation
temperatures.%
Different experiments demonstrate similar shape of dose dependence of the
elastic moduli: a rather sharp increase, reaching maximum and subsequent
decrease. Below we propose simple empirical expression for dose dependence of
the structure factor $f\left(  \Phi\right)  $.

We assume that there are two main contributions to the structure factor:
\begin{equation}
f\left(  \Phi\right)  \simeq1-\vartheta\left(  \Delta V/V\right)  +\Phi
/\Phi^{\ast} \label{ff}%
\end{equation}
due to evolution of subsystem of microcracks and due to irradiation hardening
of crystallites. It is shown in Ref. \cite{12n} (Eq.~8) that the first of
these contributions to elastic moduli is proportional to relative volume
change $\Delta V/V$. This contribution is responsible for reversal of graphite
shrinkage at high irradiation doses $\Phi$\cite{12n,16n}. Second contribution
linearly grows with the irradiation dose $\Phi$ (different powers of $\Phi$
give worse agreement with experiments). As one can see from Fig.~\ref{Doses}
this simple dependence~(\ref{ff}) describes adequately well all main features
of dose dependence of the elastic moduli at different irradiation
temperatures. Note also that the two fitting constants $\vartheta\simeq17$ and
$\Phi^{\ast}\simeq17$ almost do not depend (at noise level) on irradiation
temperature. The universality of these constants is a further argument in
support of equation~(\ref{ff}). Large value of dimensionless constant
$\vartheta$ is related to high contrast between elastic contributions of
crystallites and microcracks. Similar correlations between dimensional change
and Young's modulus structure factor $f\left(  \Phi\right)  $ were also
established for Gilsocarbon\cite{CNM-2012}.

In order to determine dose dependence of all other elastic moduli, we first
compare the Young's modulus $E_{\parallel,\perp},G_{\parallel,\perp}$ and
$\nu_{\parallel,\perp}$ for a typical Pile Grade A and single-crystal
non-irradiated graphites, see Table~\ref{T2}. \begin{table}[h]
\centering%
\begin{tabular}
[c]{c||cccc|cc}%
Pile & $E_{\parallel}^{(a)}$ & $G_{\parallel}^{(a)}$ & $E_{\perp}^{(c)}$ &
$G_{\perp}^{(c)}$ & $\nu_{\perp}$ & $\nu_{\parallel}$\\
Grade A & $9.2$ & $3.0$ & $4.6$ & $2.2$ & $0.06$ & $0.11$\\\hline
\ Mono- & $E_{\perp}^{(a)}$ & $G_{\perp}^{(a)}$ & $E_{\parallel}^{(c)}$ &
$G_{\parallel}^{(c)}$ & $\nu_{\perp}$ & $\nu_{\parallel}$\\
crystal & $909$ & $437$ & $30$ & $2.29$ & $0.04$ & $0.076$%
\end{tabular}
\caption{Young's ($E$) and shear ($G$) moduli of Pile Grade A \cite{n+1} and
single-crystal non-irradiated graphites in units of $\operatorname{GPa}%
=10^{10}\operatorname{dyn}$ and corresponding Poisson's ratios ($\nu$).
Superscripts show preferential directions of crystal axes.}%
\label{T2}%
\end{table}Due to random orientation of crystallites main contribution to all
macroscopic moduli $\bar{S}_{ij}$ comes from \textquotedblleft
large\textquotedblright\ monocrystalline moduli $s_{33}^{m}$ and $s_{44}^{m}$
(see section~\ref{MODULI} for description of orientational disorder effect).
Macroscopic compliance moduli $\bar{S}_{ij}$ are only weakly affected by
\textquotedblleft small\textquotedblright\ moduli $s_{11}^{m},s_{12}^{m}$ and
$s_{13}^{m}$ (see Table~\ref{T1} for relative values of all moduli) since
probability of finding basal planes along the direction of extrusion is small
enough. In addition, anisotropy of elastic properties of reactor graphite is
significantly smoothed out due to the presence of microcrack subsystem and
binder\cite{13n}. The above described effects are responsible for observation
of only moderate level of macroscopic orthotropy of graphite elasticity along
with sharply pronounced anisotropy on crystallite scale.

We conclude that dose dependence of the shear moduli of diffuse domain
structure (see Fig.~\ref{DIFFUSE}) can be approximated by the same structure
factor $f\left(  \Phi\right)  $ as for Young's modules~(\ref{E2}):
\begin{equation}
G_{\parallel}\left(  \Phi\right)  =G_{\perp}f\left(  \Phi\right)  ,\qquad
G_{\perp}\left(  \Phi\right)  =G_{\perp}f\left(  \Phi\right)  \label{G2}%
\end{equation}
An important consequence of Eqs.~(\ref{E2}) and~(\ref{G2}) for the Poisson's
ratios
\[
\nu_{\parallel}\left(  \Phi\right)  =const,\qquad\nu_{\perp}\left(
\Phi\right)  =const
\]
is confirmed experimentally\cite{n+1}. Using Eqs.~(\ref{E2}) and~(\ref{G2}) we
find dose dependence of all macroscopic compliance moduli:
\[
\bar{S}_{ij}\left(  \Phi\right)  =\bar{S}_{ij}f^{-1}\left(  \Phi\right)
\]
where $\bar{S}_{ij}$ are corresponding values of those moduli in absence of
irradiation, see Table~\ref{T1}.

Knowing experimental values of the moduli $\bar{S}_{ij}$ Eqs.~(\ref{s1}) can
be solved for the domain compliance moduli $s_{ij}$
\begin{equation}
s_{ij}\left(  \Phi\right)  =s_{ij}f^{-1}\left(  \Phi\right)  \label{sij}%
\end{equation}
Calculated values $s_{ij}$ in the absence of irradiation for Pile Grade A
graphite are also presented in Table~\ref{T1}. Substituting Eq.~(\ref{sij})
into Eqs.~(\ref{k1}), (\ref{k2}) and~(\ref{kV}) and using experimental value
$m_{2}\simeq0.3$ and for GR-280 graphite\cite{16n} we estimate the factors
\begin{equation}
k^{\parallel}\simeq-0.17,\quad k^{\perp}\simeq0.18,\quad k_{V}\simeq0.35
\end{equation}
in Eqs.~(\ref{eav}) and~(\ref{dV}) for relative shape changes of bulk graphite
samples. We conclude that bulk graphite sample deforms nonaffinally with
domain deformation.

Values of irradiation-induced dilatations $\left(  \Delta l/l\right)
_{\parallel}$ and $\left(  \Delta l/l\right)  _{\perp}$ can be estimated
measuring deformation of cylindrical samples (S) with diameter $8$ mm close to
the domain size $l$. Relative changes in length of such samples with
orientations coinciding ($^{\parallel}$) and orthogonal ($^{\perp}$) to the
extrusion direction are\cite{16n}:%
\begin{equation}
\left(  \Delta L/L\right)  _{S}^{\parallel,\perp}=\left(  \Delta l/l\right)
_{\parallel}+m_{2}^{\parallel,\perp}\left[  \left(  \Delta l/l\right)
_{\perp}-\left(  \Delta l/l\right)  _{\parallel}\right]  \label{par}%
\end{equation}
Here $m_{2}^{\parallel}\simeq0.3$ and $m_{2}^{\perp}\simeq0.7$ are second
moments of corresponding samples irradiated at temperature $550^{\circ}$C.

\begin{figure}
[tbh]
\begin{center}
\includegraphics[
height=1.856in,
width=2.1577in
]%
{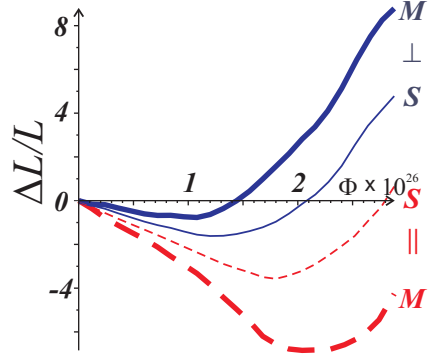}%
\caption{Dose dependence of relative deformations (in \%) of graphite
irradiated at temperature $T=550^{\circ}$C in reactor BOR-60$:$ The relative
changes of the length of cylindric finite-size samples (\emph{S}) and of
macroscopic size samples (\emph{M}) with orientations coinciding ($\parallel$,
solid lines) and orthogonal ($\perp$, dashed lines) to the extrusion
direction.}%
\label{LV}%
\end{center}
\end{figure}

Experimentally observed dose dependence of the relative length changes is
plotted in Fig.~\ref{LV}. Using these data and substituting the solution
$\left(  \Delta l/l\right)  _{\parallel,\perp}$ of Eqs.~(\ref{par}) into
Eq.~(\ref{eav}) we also plot the predicted dose dependence of relative
elongations of macroscopic samples.
Note, that relative elongation of bulk samples is much higher than that of
small size samples. In bulk samples the domains are in more crowded conditions
in directions perpendicular to the extrusion direction (in these directions
the deformation of domains due to orientational disorder is maximal).
Therefore, at low doses, the relative strain in these directions is weaker,
$|\Delta L/L|^{\perp}_{M}<|\Delta L/L|^{\perp}_{S}$. Increase in logitudinal
strain $|\Delta L/L|^{\parallel}_{M}>|\Delta L/L|^{\parallel}_{S}$ is related
to interaction between longitudinal and transverse elastic modes. These
conditions can also be rewritten in a form $(\Delta L/L)^{\perp}_{M}>(\Delta
L/L)^{\perp}_{S}$ and $(\Delta L/L)^{\parallel}_{M}<(\Delta L/L)^{\parallel
}_{S}$ that is also valid for large doses. Since obtained results are
sensitive to value of the moment $m_{2}$, it is desirable to carry out further
experiments to get better characterization of reactor grade graphites texture.

\section{Summary}

The main results of this paper are as follows:

1. Basing on experimental data on radiation-induced deformation of graphite,
we introduced the concept of diffuse domain structure, which is the reason of
size effects in reactor graphite. In such structure there are no sharp
boundaries between domains, and the direction of domain anisotropy undergoes
random deviations from global orientation of the sample. We also derived
equations of macroscopic elasticity, describing the deformation of solids due
to irradiation-induced shape change of their domains having transversely
isotropic structure. This theory can also be used to describe deformation of
stressed samples. A relation is obtained between elastic moduli of domains and
macroscopic solid as the whole. We have proposed a scheme for conversion of
experimental data obtained on samples of finite size to describe shape-change
of bulk graphite. Such scheme is required for engineering calculations of
graphite blocks integrity.

2. We have presented simple model explaining origin of diffuse domains in
graphite produced by the extrusion process. Our estimation of domain size is
in good agreement with known experimental data\cite{16n}. This model can also
be used to predict the dependence of the domain size on extrusion velocity and
viscosity of visco-plastic medium.

3. We also elaborated simple model describing orientational ordering of
domains during extrusion. This model enables us to calculate the moments of
domain's orientation, needed in description of elastic properties of
reactor-grade graphite.


\begin{thebibliography}{99}                                                                                               %


\bibitem {1n}John H.W. Simmons, Radiation damage in graphite, Oxford, Pergamon
Press (1965).

\bibitem {2n}B.T. Kelly, W.H. Martin and P.T. Nettley, \emph{Phyl. Trans. A},
\textbf{260} (1966).

\bibitem {3n}B.T. Kelly, \emph{Prog. Nucl. Energy}, \textbf{2} (1978) 219-269.

\bibitem {4n}Irradiation damage in graphite due to fast neutrons in fission
and fusion systems, IAEA-TECDOC-1154, September (2000).

\bibitem {5n}G. Haag, Properties of ATR-2E Graphite and property changes due
to fast neutron irradiation, Forschungszentrum, J\"{u}lich, (2005).

\bibitem {6n}Ya.I. Shtrombakh, B.A. Gurovich, P.A. Platonov and V.M. Alekseev,
\emph{J. of Nucl. Mater.,} \textbf{225} (1995) 273-301.

\bibitem {7n}G. Hall, B.J. Marsden, S.L. Fok, J. Smart, \emph{Nucl. Engin. and
Design}, \textbf{222} (2003) 39-330.

\bibitem {21n}J.F. Nye, Physical properties of crystals, Oxford, Clarendon
Press, (1964).

\bibitem {8n}D.K.L. Tsang, B.J. Marsden, \emph{J. of Nucl. Mater.,}
\textbf{350} (2006) 208-220.

\bibitem {9n}G. Hall, B.J. Marsden, S.L. Fok, \emph{J. of Nucl. Mater.,}
\textbf{353} (2006) 12-18.

\bibitem {10n}C. Berre et al. \emph{J. of Nucl. Mater.,} \textbf{352} (2006) 1-5.

\bibitem {11n}T.D. Burchell, Proc. IAEA Specialist's Meeting on the Present
Status of Graphite Development for Gas-Cooled Reactors, Sept. 9 (1991) 49-58.

\bibitem {n+7}A.V. Subbotin, O.V. Ivanov, I.M. Dremin et al. \emph{Atomic
Energy}, \textbf{100} (2005) 204-26.

\bibitem {12n}S.V. Panyukov and A.V. Subbotin, \emph{Atomic Energy},
\textbf{105} (2008) 25-32.

\bibitem {13n}S.V. Panyukov and A.V. Subbotin, \emph{Atomic Energy},
\textbf{107} (2009) 268-273.

\bibitem {15n}W.N. Reynolds and P.A. Thrower, \emph{Phyl. Mag.} \textbf{12}
(1965) 573-593.

\bibitem {13a}S. Amelinckx, P. Delavignette and M. Heerschap, Dislocations and
stacking faults in graphite, \emph{Chem. Phys. Carbon,} ed. P.L. Walker 1
(1965) 1-71.

\bibitem {15a}T.D. Burchell and T. Oku, Material properties data for fusion
reactor plasma facing carbon-carbon composites Atomic and Plasma--Material
Interaction Data for Fusion (supplement to J. Nucl. Fusion) 5 (1994) 77-128
(International Atomic Energy Agency).

\bibitem {14n}W.C. Morgan, \emph{Carbon}, \textbf{1} (1964) 255-261, 51-62.

\bibitem {16n}M.V. Arjakov, A.V. Subbotin, S.V. Panyukov et al, \emph{J. of
Nucl. Mater.,} \textbf{420} (2012) 241-251.

\bibitem {17n}R.E. Nightingale and E.M. Woodruff, \emph{Nucl. Sci. and
Engineering,} \textbf{19} (1964) 390-392.

\bibitem {18n}A.E.H. Love, Mathematical theory of elasticity, London, (1952).

\bibitem {19n}S.G. Lekhnitsky, Elastic theory of anisotropic medium, Nauka,
Moscow, 1977 (in Russian).

\bibitem {20n}C. Baker and A. Kelly, \emph{Phil. Mag.} \textbf{9} (1964), 927.

\bibitem {Voigt-89}W. Voigt, \emph{Wied. Ann.} \textbf{38} (1889) 573--587.

\bibitem {Reuss-29}A. Reuss, \emph{Z. Angew. Math. Mech.} \textbf{9} (1929) 49--58.

\bibitem {CNM-2012}B.J. Marsden, G.N. Hall, \emph{Compr. Nucl. Mat.},
\textbf{4} (2012) 325-390.

\bibitem {n+1}Irradiation damage in graphite due to fast neutrons in fission
and fusion systems, IAEA-TECDOC-1154, September (2000).

\bibitem {n+4}E.J. Seldin and C.W. Nezbeda, \emph{J. Appl. Phys. }\textbf{41}
(1970) 3389.

\bibitem {n+6}J.B. Ayasse and E. Bonjour, Pro. fourth SCI Conference on
industrial carbons and graphites, SCI, London, (1976).
\end{thebibliography}
\end{document}